\documentclass[]{IEEEtran}
\IEEEoverridecommandlockouts
\ifCLASSINFOpdf
\else
\fi
\usepackage[cmex10]{amsmath}
\usepackage{array}

\usepackage{mdwmath}
\usepackage{mdwtab}
\usepackage{fixltx2e}
\usepackage{cite}
\usepackage{graphicx}
\usepackage{amssymb}
\usepackage{amsmath}
\usepackage{amsthm}
\usepackage{amsbsy} 
\usepackage{float}
\usepackage{balance}
\usepackage{url}

\usepackage{multicol}
\usepackage{float}
\newfloat{longequation}{b}{ext}

\newfloat{eq.}{t}{ext}

\newtheorem{thm}{Theorem}[section]

\newtheorem{lem}{Lemma}[section]
\newtheorem{pro}{Property}[section]
\newtheorem{df}{Definition}[section]
\newtheorem{rem}{Remark}[section]
\usepackage{color}

\begin{document}

%
\title{Large System Analysis for \\Amplify \& Forward
SIMO Multiple Access Channel
with Ill-conditioned Second Hop}
%
%
%

\author{Symeon~Chatzinotas
\thanks{S. Chatzinotas is with the Interdisciplinary Centre for Security, Reliability and Trust, University of Luxembourg  (http://www.securityandtrust.lu)
 e-mail: Symeon.Chatzinotas@uni.lu.}
}

\maketitle

\begin{abstract}
Relaying has been extensively studied during the
last decades and has found numerous applications in wireless
communications. The simplest relaying method, namely amplify
and forward, has shown potential in MIMO multiple access
systems, when Gaussian fading channels are assumed for both
hops. However, in some cases ill conditioned channels may appear
on the second hop. For example, this impairment could affect
cooperative BS systems with microwave link backhauling, which
involve strong line of sight channels with insufficient scattering.
In this paper, we consider a large system analysis of such as
model focusing on both optimal joint decoding and joint MMSE filtering receivers.
Analytical methods based on free probability are presented
for calculating the ergodic throughput, the MMSE error and the average SINR. Furthermore, the
performance degradation of the system throughput is evaluated considering second
hop impairments such as ill-conditioning and rank deficiency, while high- and low-SNR limits are calculated for the considered performance metrics. Finally, the cooperative BS system is compared to a conventional channel resource
division strategy and suitable operating points are proposed.  \end{abstract}


%
\IEEEpeerreviewmaketitle

\section{Introduction}

The Dual Hop (DH) Amplify-and-Forward (AF)  relay channel has attracted a great deal of attention mainly due to its low complexity and its manyfold benefits, such as coverage extension and decreased outage probability. Although the DH AF channel has been extensively studied in the literature \cite{Morgenshtern2007,Jin2010,Wagner08}, the effect of second hop condition number on its performance is not well quantified yet. 

Assuming Gaussian channel matrices in both hops, authors in \cite{Morgenshtern2007} approached the problem asymptotically using Silverstein's fixed-point equation and found closed-forms expressions for the Stieltjes transform. Under similar assumptions, a finite analysis was recently performed by \cite{Jin2010}. On the other hand, authors in \cite{Wagner08}  following a replica analysis tackled the problem of Kronecker correlated Gaussian matrices. 

In addition, the MIMO MAC has been studied heavily during the last decades since it comprises a fundamental channel model
for multiuser uplink cellular \cite{Chatzinotas_JWCOM} and multibeam return link communications\cite{Letzepis2008,Christopoulos2013}. The work in \cite{Chatzinotas_WCSP2012,Chatzinotas_WCL2013} has combined AF relaying with a MAC  and has performed a free-probabilistic analysis for channel capacity. Furthermore, the work in \cite{Wen2010} has combined AF relaying with cooperative Base Stations  and has performed a replica analysis for channel capacity and MMSE throughput.  

In our scenario, we study a DH AF SIMO MAC modelling cooperative BSs with microwave link backhauling and we focus on the impact of ill-conditioned or rank-deficient MIMO channel matrices in the second hop. The paradigm of BS cooperation (also known as multicell joint decoding and network MIMO) was initially proposed almost three decades ago and its  performance gain over conventional cellular systems was demonstrated in two seminal papers \cite{Hanly1993,Wyner1994}. The main assumption is the existence of a central processor (CP) which is interconnected to all the BSs through a backhaul of wideband, delayless and error-free links. In addition, the central processor is assumed to have perfect Channel State Information (CSI) about all the wireless links of the system. These assumptions enable the central processor to jointly decode all the UTs of the system, rendering the concept of intercell interference void. Since then, there has been an ongoing research activity extending and modifying the initial results for more practical propagation environments, transmission techniques and backhaul infrastructures in an attempt to better quantify the performance gain. 

More specifically, it was demonstrated in \cite{Somekh2000} that Rayleigh fading promotes multiuser diversity which is beneficial for the ergodic capacity performance. Subsequently, realistic path-loss models and user distribution were investigated in \cite{Chatzinotas_letter,Chatzinotas_SPAWC} providing closed-form capacity expressions based on the cell size, path loss exponent and user spatial p.d.f. The beneficial effect of MIMO links was established in \cite{Aktas2006,Chatzinotas_ISWCS}, where a linear scaling with the number of BS antennas was proven. However, correlation between multiple antennas has an adverse effect as shown in \cite{Chatzinotas_JWCOM}, especially when correlation affects the BS-side. 

Regarding backhauling, the ideal assumptions of previous studies can only be satisfied by fiber connectivity between all BSs and the central processor. However, in current backhaul infrastructure microwave links are often used, especially in rural environments where the cable network is unavailable.  Recent studies have tried to alleviate the perfect backhaul assumption by focusing in finite-rate errorless links to the CP \cite{Sanderovich2009}, finite-rate errorless links between adjacent BSs \cite{Simeone2009} and finite-sum-rate backhaul with imperfect CSI \cite{Marsch2011}. Contrary to these approaches, this paper assumes microwave backhauling from all BSs to the CP, operating over the same frequency. The BSs amplify and forward the received signals to an antenna array at the CP and thus the backhaul rate is limited by the system geometry, the relaying power and the impairments of the second hop MIMO channel. 

In this direction, the main contributions of this paper are:
\begin{itemize}
\item 
the derivation of the ergodic capacity 
and a lower  bound
on the average Minimum Mean Square Error (MMSE)
for AF
SIMO MAC
with Ill-conditioned second hop
\item
the derivation of high and low SNR limits for channel capacity and MMSE performance 
\item
the evaluation of the condition number and normalized rank of the second hop channel matrix  on the system performance
\item
the performance comparison to a conventional system which employs resource division access to eliminate multiuser interference. 
\end{itemize}

The remainder of this paper is structured as follows: Section \ref{sec: system model} introduces the system model, while section \ref{sec: performance analysis} describes the free probability derivations and the main capacity and MMSE results. Section \ref{sec: numerical results} verifies the accuracy of the analysis by comparing with Monte Carlo simulations and evaluates the effect of various system parameters on the performance. Section \ref{sec: conclusion} concludes the paper.

\subsection{Notation}
Throughout the formulations of this paper, normal \(x\), lower-case boldface \(\mathbf x\) and upper-case boldface \(\mathbf X\) font is used for scalars, vectors and matrices respectively. \(\mathbb{E}[\cdot]\) denotes the expectation,  \(\left(\cdot\right)^H\) denotes the conjugate transpose matrix, and \(\odot\) denotes the Hadamard product. The Frobenius norm of a matrix or vector is denoted by \(\left\Vert \cdot\right\Vert\), the absolute value of a scalar is denoted by \(\left\vert \cdot\right\vert\) and the delta function is denoted by 
$\delta(\cdot)$. $(\cdot)^+$ is equivalent to \(\max(0,\cdot), 1\left\{ \cdot \right\}\) is the indicator function and \(\rightarrow\) denotes almost sure (a.s.) convergence.

\section{System Model}
\label{sec: system model}

\subsection{Input-output Model}
\label{subsec: Application Scenario}
\begin{figure}
        \centering
                \includegraphics[width=0.48\textwidth]{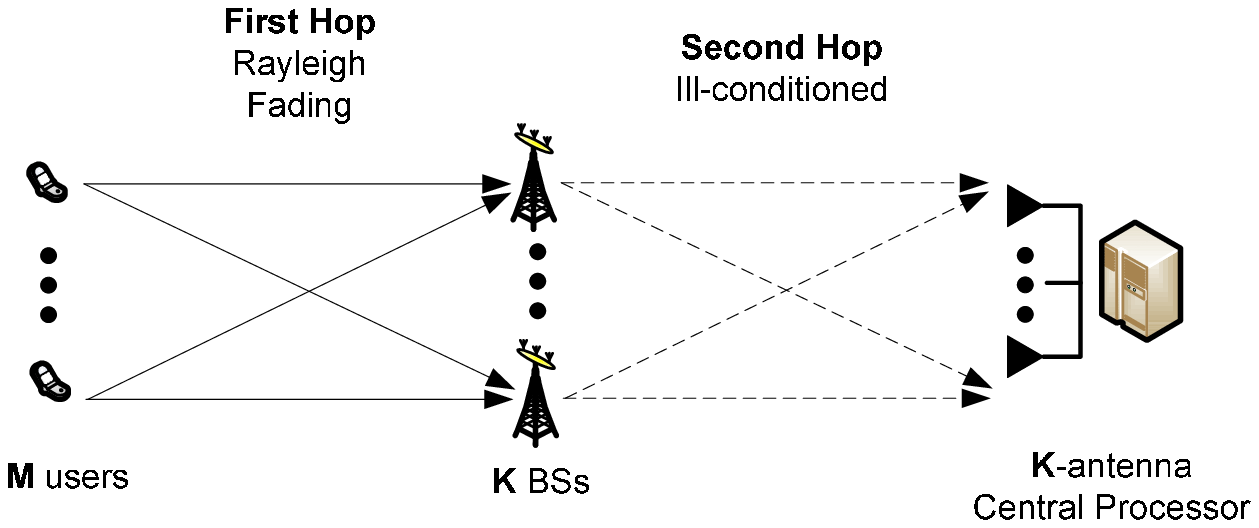}
        \caption{Conceptional illustration of the system model.}
        \label{fig: block}
\end{figure}
Figure \ref{fig: block} is a conceptual illustration of the input-output model, which included \(M\) users, \(K\) BSs and a CP equipped with a \(K\)-antenna array. It can be seen that the BS-CP (Central Processor) microwave links (second hop) form an ill-conditioned SIMO MAC, whereas the user-BS-CP  links can be modelled as SIMO AF MAC. Gaussian input is considered at the user-side, while neither users nor relays are aware of the Channel State Information (CSI). On the other hand, the CP is assumed to have perfect knowledge of  system-wide CSI. The described channel model can be expressed as follows:
\begin{align}
\mathbf{y}_1&=\mathbf{H}_1\mathbf{x}_1+\mathbf{z}_1\nonumber
\\\mathbf{y}_2&=\mathbf{H}_2\sqrt\nu\mathbf{y}_1+\mathbf{z}_2\Leftrightarrow\nonumber
\\\mathbf{y}_2&=\sqrt\nu\mathbf{H}_2\mathbf{H}_1\mathbf{x}_1+\sqrt\nu\mathbf{H}_2\mathbf{z}_1+\mathbf{z}_2,
\label{eq: generic channel model}
\end{align}
where the $M\times 1$ vector $\mathbf{x}_1$ denotes the  user transmitted symbol vector with individual Signal to Noise Ratio (SNR) \(\mu\) ($\mathbb{E}[\mathbf{x}_1\mathbf{x}_1^H]=\mu\mathbf{I}$), $\mathbf{y}_1$ denotes the $K\times 1$ received symbol vector by the BSs and the $K\times 1$ vector $\mathbf{z}_1$ denotes AWGN at BS-side with $\mathbb{E}[\mathbf{z}_1]=\mathbf{0}$ and $\mathbb{E}[\mathbf{z}_1\mathbf{z}_1^H]=\mathbf{I}$. The received signal $\mathbf{y}_1$ is amplified by $\nu$ and forwarded and as a result $\mathbf{y}_2$ denotes the $K\times 1$ received symbol vector by the CP and the $K\times 1$ vector $\mathbf{z}_2$ denotes AWGN at CP-side with $\mathbb{E}[\mathbf{z}_2]=\mathbf{0}$ and $\mathbb{E}[\mathbf{z}_2\mathbf{z}_2^H]=\mathbf{I}$. It should be noted that for the remainder of this document \(\mu\) and \(\nu\) will be referred to as First Hop Power (FHP) and Second Hop Power (SHP) respectively.

The $K\times M$ channel matrix $\mathbf{H}_1$ and the $K\times K$ channel matrix $\mathbf{H}_2$ represent the concatenated channel vectors for the user-BS and BS-CP links respectively.
The first hop Rayleigh
fading channel $\mathbf{H}_1\sim\mathcal{C}\mathcal{N}(\mathbf{0,I})$ can be modelled as a Gaussian matrix with independent identically distributed (i.i.d.) complex circularly symmetric (c.c.s.) elements. The BSs-CP channel \(\mathbf{H}_2\) under line of sight suffers from
correlation due to lack of scattering and thus it can be modelled as an ill-conditioned deterministic channel with variable condition number \(\zeta^2=\lambda_{\max}(\mathbf{H}_2\mathbf{H}_2^H)/\lambda_{\min}(\mathbf{H}_2\mathbf{H}_2^H)\) or as a rank-deficient deterministic channel with variable normalized rank \(\alpha=\mathrm{rank(\mathbf{H}_2\mathbf{H}_2^H)}/K\). The exact matrix models  for \(\mathbf{H}_2\) are described in detail in sections \ref{sec: Ill-Conditioned Second Hop} and \ref{sec: rank-deficient second hop}.
\subsection{Performance Metrics}
The performance metrics considered in this work are the channel capacity achieved by successive interference cancellation at the CP and the average Mimimum Mean Square Error  (MMSE) achieved by joint MMSE filtering at the CP followed by single-user decoding. It should be noted that both of these receiver structures require multiuser processing at the CP. On the other hand, section  \ref{subsec: conventional} considers a conventional system where Frequency or Time Division Multiple Access is used in combination with single-user interference-free decoding at the CP. 

The capacity per receive antenna of this channel model is given by  \cite{Foschini1998,Blum2003,Lozano2002,ChatzinotasElsevier11}:
\begin{align}
\mathrm{C}&=\frac{1}{K}\mathbb{E}\left[\log\det\left(\mathbf{I}+\mu\nu\mathbf{H}_2\mathbf{H}_1\mathbf{H}_1^H\mathbf{H}_2^H\left(\mathbf{I}+\nu\mathbf{H}_{2}\mathbf{H}_{2}^H\right)^{-1}\right)\right]\label{eq: first form}
\\&=\frac{1}{K}\mathbb{E}\left[\log\det\left(\mathbf{I}+\nu\mathbf{H}_{2}\mathbf{H}_{2}^H+\mu\nu\mathbf{H}_2\mathbf{H}_1\mathbf{H}_1^H\mathbf{H}_2^H\right)\right]\nonumber\\&-\frac{1}{K}\mathbb{E}\left[\log\det\left(\mathbf{I}+\nu\mathbf{H}_{2}\mathbf{H}_{2}^H\right)\right]
\label{eq: original capacity}
\\&\stackrel{_{(a)}}{=}\frac{1}{K}\mathbb{E}\left[\log\det\left(\mathbf{I}+\nu\mathbf{H}_{2}^H\mathbf{H}_{2}\left(\mathbf{I}+{\mu}\mathbf{H}_{1}\mathbf{H}_{1}^H\right)\right)\right]\nonumber\\&-\frac{1}{K}\mathbb{E}\left[\log\det\left(\mathbf{I}+\nu\mathbf{H}_{2}^H\mathbf{H}_{2}\right)\right]=\mathrm{C}_1-\mathrm{C}_2
\label{eq: generic capacity}
\end{align}
where step \((a)\) uses the property \(\log\det(\mathbf{I+AB})=\log\det(\mathbf{I+BA})\).
It can be observed that the positive term $\mathrm{C}_1$ corresponds to the mutual information due to relaying, while the negative term $\mathrm{C}_2$ represents the performance loss due to noise amplification.

The receiver complexity in order to achieve the channel capacity is quite high since it involves successive interference cancellation \cite{Chatzinotas_IWCMC}. In this direction, we consider a less complex receiver which involves  multiuser MMSE filtering followed by single-user decoding. Since, this is a linear operation we assume that \(K=M\). The performance of the  MMSE receiver is dependent on the achieved MSE averaged over users and channel realizations and is given by: 
\begin{align}
&\mathrm{mmse_{avg}}=\mathbb{E}\left[\frac{1}{M}\sum_{m=1}^M\mathrm{mmse}_m\right]\nonumber
\\=\mathbb{E}&\left[\frac{1}{M}\sum_{m=1}^M\left[ \left(\mathbf{I}+\mu\nu\mathbf{H}_1^H\mathbf{H}_2^H\left(\mathbf{I}+\nu\mathbf{H}_{2}\mathbf{H}_{2}^H\right)^{-1}\mathbf{H}_2\mathbf{H}_1\right)^{-1} \right]_{m,m}\right]\nonumber
\\=\mathbb{E}&\left[\frac{1}{M}\mathrm{tr}\left\{ \left(\mathbf{I}+\mu\nu\mathbf{H}_1^H\mathbf{H}_2^H\left(\mathbf{I}+\nu\mathbf{H}_{2}\mathbf{H}_{2}^H\right)^{-1}\mathbf{H}_2\mathbf{H}_1\right)^{-1} \right\}\right]\nonumber
\\=\mathbb{E}&\left[\frac{1}{M}\mathrm{tr}\left\{ \left(\mathbf{I}+\nu\mathbf{H}_{2}\left(\mathbf{I}+{\mu}\mathbf{H}_{1}\mathbf{H}_{1}^H\right)\mathbf{H}_{2}^H\right)^{-1}\left(\mathbf{I}+\nu\mathbf{H}_{2}\mathbf{H}_{2}^H\right) \right\}\right].
\label{eq: mmse avg simulation}
\end{align}

The average SINR and the achieved throughput per receive antenna using LMMSE is given by: 
\begin{align}
&\mathrm{SINR_{avg}}=\mathbb{E}\left[\frac{1}{M}\sum_{m=1}^M\mathrm{mmse}_m^{-1}\right]-1
\\&\mathrm{C_{mmse}}=\log\left( 1+\mathrm{SINR_{avg}}\right)\geq-\log\left( \mathrm{mmse_{avg}}\right)\nonumber
\end{align}
\begin{scriptsize}
\begin{align}
=-\log\left( \frac{1}{M}\mathbb{E}\left[ \mathrm{tr}\left\{ \left(\mathbf{I}+\nu\mathbf{H}_{2}\left(\mathbf{I}+{\mu}\mathbf{H}_{1}\mathbf{H}_{1}^H\right)\mathbf{H}_{2}^H\right)^{-1}\left(\mathbf{I}+\nu\mathbf{H}_{2}\mathbf{H}_{2}^H\right) \right\} \right]\ \right).
\label{eq: mmse capacity simulation}
\end{align}
\end{scriptsize}
Compared to existing literature, our work starts from eq. \eqref{eq: generic capacity} since the original problem in eq. \eqref{eq: original capacity} yields quite involved solutions\cite{Morgenshtern2007,Jin2010,Wagner08}. 
In addition, by decomposing the problem in two components, deeper insights can be acquired. We follow a free probabilistic analysis as in \cite{Muller2002,Chatzinotas_JWCOM,ChatzinotasElsevier11,ChatzinotasEUSIP11,Sharma2013} to derive the channel capacity, but we extend it for the described DH AF SIMO MAC including the noise amplification terms and ill-conditioned second hop modelling. More importantly, we consider the MMSE filtering receiver and we obtain a lower bound on the average MMSE performance.

To simplify the notations during the mathematical analysis, the following auxiliary variables are defined:
\begin{align}
\mathbf{M}&=\mathbf{I}+{\mu}\mathbf{H}_{1}\mathbf{H}_{1}^H\nonumber\\
\mathbf{\tilde M}&=\mathbf{I}+{\nu}\mathbf{H}_{2}\mathbf{H}_{2}^H\nonumber\\
\mathbf{{N}}&=\mathbf{H}_{1}\mathbf{H}_{1}^H\nonumber\\
\mathbf{\tilde N}&=\mathbf{H}_2^H\mathbf{H}_2\nonumber\\
\mathbf{K}&=\mathbf{H}_2^H\mathbf{H}_2\left(\mathbf{I}+{\mu}\mathbf{H}_{1}\mathbf{H}_{1}^H\right)=\mathbf{\tilde NM}\nonumber\\
\mathbf{\tilde K}&=\mathbf{H}_2\left(\mathbf{I}+{\mu}\mathbf{H}_{1}\mathbf{H}_{1}^H\right)\mathbf{H}_2^H\nonumber\\
\beta&=\frac{M}{K}\nonumber
\end{align}
where $\beta\geq1$ is the ratio of horizontal to vertical dimensions of matrix $\mathbf{H}_1$ (users/BS). 
\subsection{Conventional System}
\label{subsec: conventional}
In a conventional cellular system, the available resources (frequency or time) would have to be split in \(K\) pieces in order to avoid multiuser interference from neighboring BSs. This entails that only \(K\) out of \(M\) users could be served simultaneously, namely one user per cell (\(\beta=1\)). Moreover, this is the usual approach employed by current standards in order to avoid co-channel interference\footnote{In reality, higher frequency reuse can be used in order to exploit spatial separation of cells. However, frequency reuse cannot be exploited in the considered system without creating multiuser interference in the CP through the AF relaying.}. On the plus side, each user or BS relay could concentrate its power on a smaller portion of the resource using \(K\mu\) and \(K\nu\) respectively. Assuming a single user per cell (\(K=M\)), the conventional channel model for a single user-BS-CP link can be written as:
\begin{align}
{y}_1&={h}_1{x}_1+{z}_1\nonumber
\\{y}_2&={h}_2\sqrt{K\nu}{y}_1+{z}_2\Leftrightarrow\nonumber
\\{y}_2&=\sqrt{K\nu}{h}_2{h}_1{x}_1+\sqrt{K\nu}{h}_2{z}_1+{z}_2
\label{eq: generic conventional channel model}
\end{align}
with \(x_1\) Gaussian input with \(\mathbb{E}[x_1^2]=K\mu\) and $z_1,z_2$ AWGN with  \(\mathbb{E}[z_1^2]=\mathbb{E}[z_2^2]=1\).
In this case, the per-antenna capacity at the CP would be:
\begin{equation}
\mathrm{C_{co}}=\mathbb{E}\left[\log\left( 1+\mathrm{SNR} \right)\right]=\mathbb{E}\left[\log\left( 1+\frac{K^2\nu{h_2}^2 \mu{h_1}^{2}}{1+K\nu {h_2}^2} \right)\right],
\end{equation}
where \(h_1\) and \(h_2\) are the channel coefficients of the first and second hop respectively. The first and second hop are modelled as  Rayleigh fading and AWGN channels respectively and thus we can assume that \(h_{1}\sim\mathcal{C}\mathcal{N}(0,1)\) and \(h_2=1\). The performance of the conventional and proposed transmission schemes are compared in section \ref{subsec: comparison}. 

\section{Performance Analysis}
\label{sec: performance analysis}

In order to calculate the system performance analytically, we resort to asymptotic analysis which entails that the dimensions of the channel matrices grow to infinity assuming proper normalizations. It has already been shown in many occasions that asymptotic analysis yields results which are also valid for finite dimensions \cite{Lozano2002,Tulino04,Martin2004}. In other words, the expressions of interest converge quickly to a deterministic value as the number of channel matrix dimensions increases. 

In this direction, the components of eq. \eqref{eq: generic capacity} can be written asymptotically as: 
\begin{align}
\mathrm{C}_1&=\frac{1}{K}\lim_{K,M\rightarrow\infty}\mathbb{E}\left[\log\det\left(\mathbf{I}+\nu\mathbf{H}_2^H\mathbf{H}_2\left(\mathbf{I}+\mu\mathbf{H}_\mathrm{1}\mathbf{H}_\mathrm{1}^H\right)\right)\right]\nonumber\\
&=\lim_{K,M\rightarrow\infty}\mathbb{E}\left[\frac{1}{K}\sum_{i=1}^K \log\left(1+\nu\lambda_i\left(\mathbf{K}\right)\right)\right]\nonumber\\
&\rightarrow\int_0^\infty \log\left(1+\nu x\right)f^\infty_{\mathbf{K}}\left(x\right)\mathrm{d}x,
\label{eq: capacity aepdf 1}
\end{align}
\begin{align}
\mathrm{C}_2&=\frac{1}{K}\lim_{K,N\rightarrow\infty}\mathbb{E}\left[\log\det\left(\mathbf{I}+\nu\mathbf{H}_2^H\mathbf{H}_2\right)\right]\nonumber\\
&=\lim_{K,N\rightarrow\infty}\mathbb{E}\left[\frac{1}{K}\sum_{i=1}^K \log\left(1+\nu\lambda_i\left(\mathbf{\tilde N}\right)\right)\right]\nonumber\\
&\rightarrow\int_0^\infty \log\left(1+\nu x\right)f^\infty_{\mathbf{\tilde N}}\left(x\right)\mathrm{d}x,
\label{eq: capacity aepdf 2}
\end{align}
where $\lambda_i\left(\mathbf{X}\right)$ is the $i$th ordered eigenvalue of matrix $\mathbf{X}$ and $f^\infty_{\mathbf{X}}$ is the asymptotic eigenvalue probability density function (a.e.p.d.f.) of $\mathbf{X}$.
It should be noted that while the channel dimensions $K,M$ grow to infinity, the matrix dimension ratio  $\beta$ is kept constant. 

Using a similar approach, the average MMSE when \(\beta=1\) can be expressed as:
\begin{align}
\mathrm{mmse_{avg}}
=&\lim_{K,M\rightarrow\infty}\mathbb{E}\left[\frac{1}{M}\mathrm{tr}\left\{ \left(\mathbf I+\nu\mathbf{\tilde K}\right)^{-1}\mathbf{\tilde M} \right\}\right]
\\\stackrel{_{(a)}}{\geq}&\lim_{K,M\rightarrow\infty}\mathbb{E}\left[\frac{1}{M}\sum_{m=1}^M\frac{\lambda_{M-m+1}\left(\mathbf{\tilde M}\right)}{1+\nu\lambda_m \left(\mathbf{\tilde K}\right)}\right]\nonumber
\\\rightarrow&\int_0^1\frac{F^{-1}_{\mathbf{\tilde M}}(1-x)}{1+\nu F^{-1}_{\mathbf{\tilde K}}(x)} \mathrm dx
\label{eq: average MMSE}
\end{align}
where step \((a)\) follows from property $\mathrm{tr}\{\mathbf{AB}\}\geq \sum_{m=1}^M \lambda_m(\mathbf A)\lambda_{M-m+1}(\mathbf B)$ in  \cite{Lasserre1995} and \(F^{-1}_{\mathbf{X}}\) denotes the inverse function of the asymptotic eigenvalue cumulative density function (a.e.c.d.f.). The last step follows from the fact that the ordered eigenvalues can be obtained by uniformly sampling the inverse c.d.f. in the asymptotic regime \cite{Letzepis2008}.


To calculate the expression of eq. \eqref{eq: capacity aepdf 1},\eqref{eq: capacity aepdf 2},\eqref{eq: average MMSE}, it suffices to derive the asymptotic densities of $\mathbf{K},\mathbf{\tilde N},\mathbf{\tilde K},\mathbf{\tilde M}$, which can be achieved through the principles of free probability theory \cite{Voiculescu83,Hiai2000,Hiai00,Bai1999} as described in sections \ref{subsec: fading hop} and \ref{sec: Ill-Conditioned Second Hop}.
Free probability (FP) has been proposed by Voiculescu \cite{Voiculescu83} and has found numerous applications in the field of wireless communications. More specifically, FP\ has been applied for capacity derivations of variance profiled \cite{Chatzinotas_letter}, correlated \cite{Chatzinotas_JWCOM} Rayleigh channels, as well as Rayleigh product channels \cite{Muller2002}. Furthermore, it has been used for studying cooperative relays \cite{Li2008}, interference channels \cite{ChatzinotasElsevier11} and interference alignment scenarios \cite{ChatzinotasEUSIP11}.
The advantage of FP methodology compared to other techniques, such as Stieltjes method, replica analysis and deterministic equivalents, is that the derived formulas usually require just a polynomial solution instead of fixed-point equations. However, the condition for these simple solutions is that the original aepdfs can be expressed in polynomial form \cite{Letzepis2007}. For completeness, some preliminaries of Random Matrix Theory have been included in appendix \ref{ap: preliminaries} in order to facilitate the comprehension of derivations in sections \ref{subsec: fading hop}, \ref{sec: Ill-Conditioned Second Hop} and \ref{sec: rank-deficient second hop}.
\subsection{Fading First Hop}
\label{subsec: fading hop}
The first hop from users to BSs can
be modelled as a Rayleigh fading channel, namely $\mathbf{H}_1\sim\mathcal{C}\mathcal{N}(\mathbf{0,I})$\footnote{This analysis can be straightforwardly extended for cases where variable received power is considered for each BS due to variable transmit powers or propagation paths across users. In this case, the channel can be modeled as a variance-profiled Gaussian  matrix and it can be tackled using a scaling approximation as described in \cite{Chatzinotas_letter,Chatzinotas_JWCOM}. }.
 \begin{df}
\label{df: MP law}
Considering a Gaussian \(K\times M\) channel matrix \(\mathbf{H}_1\sim\mathcal{C}\mathcal{N}\left( \mathbf{0},\mathbf{I} \right)\), the a.e.p.d.f. of \(\frac{1}{K}\mathbf{H}_1\mathbf{H}_1^H\)
converges almost surely (a.s.) to the non-random limiting eigenvalue distribution
of the Mar\v cenko-Pastur law \cite{Marcenko1967}, whose density functions are given
by
\begin{align}
        f^\infty_{\frac{1}{K}\mathbf{H}_1\mathbf{H}_1^H}(x)&{\rightarrow}\  f_{\mathrm{MP}}(x,\beta)
        \nonumber
\\f_{\mathrm{MP}}\left(x,\beta\right)&=\left(1-\beta\right)^+\delta\left(x\right)+\frac{\sqrt{\left(x-a\right)^+\left(b-x\right)^+}}{2\pi x}\label{eq: MP aepdf}
\end{align}
where $a=(1-\sqrt{\beta})^2,b=(1+\sqrt{\beta})^2$
and \(\eta\)-transform,  $\Sigma$-transform and Shannon transform are given by \cite{Tulino04}
\begin{align}
\label{eq: n transform of Marcekno Pastur law}
\eta_{\mathrm{MP}}\left(x,\beta \right)&=1-\frac{\phi\left(x,\beta
\right)}{4 x}
\\\phi\left( x,\beta\right)=&\left(\sqrt{x\left( 1+\sqrt{\beta} \right)^2+1}-\sqrt{x\left( 1-\sqrt{\beta}\right)^2+1}\right)^{2}\nonumber
\\\Sigma_{\mathrm{MP}}\left(x,\beta \right)&=\frac{1}{\beta+x}\label{eq: sigma transform N}
\\\mathcal{V}_{\mathrm{MP}}\left(x,\beta\right)&=\beta\mathrm{log}\left( 1+x-\frac{1}{4}\phi\left( x,\beta \right)\ \right)\nonumber
\\&+\mathrm{log}\left( 1+x \beta-\frac{1}{4}\phi\left( x,\beta \right)\ \right) -\frac{1}{4x}\phi\left( x,\beta\right).
\end{align}

\end{df}

\begin{lem}
The cumulative density function of the Mar\v cenko-Pastur law for \(\beta=1\) is given by:
\begin{equation}
F_{\mathrm{MP}}\left(x\right)={\frac { \sqrt {-x \left( x-4 \right) }+2\arcsin \left( -1+ x/2
 \right)+\pi }{2\pi }}.
\end{equation}
\begin{proof}
The c.d.f. follow from eq.\eqref{eq: MP aepdf} after integration for \(\beta=1\).
\end{proof}
\end{lem}

\begin{lem}
\label{lem: M aepdf}
The a.e.p.d.f. of \(\mathbf M\)
converges almost surely (a.s.) to:

\begin{scriptsize}
\begin{equation}
        f^\infty_{\mathbf M}(x,\beta,\bar \mu){\rightarrow}  {\frac {\sqrt { \left( x-1-\bar \mu+2\bar  \mu\sqrt {\beta}-\bar \mu \beta \right) 
 \left( \bar \mu+2\bar  \mu\sqrt {\beta}+\bar \mu \beta-x+1 \right) }}{2\bar \mu\pi   \left( x-1
 \right) }},
 \label{eq: I plus MP aepdf}
\end{equation}
\end{scriptsize}
where \(\bar \mu=K\mu.\)
\begin{proof}
The a.e.p.d.f. can be calculated considering the transformation  $z(x)=(1+K\mu\ x)$, where $z$ and $x$ represent the eigenvalues of $\mathbf M$ and $\frac{1}{K}\mathbf{H}_1\mathbf{H}_1^H$ respectively:
\begin{align}
f^\infty_{\mathbf M}(x)&=\left| \frac{1}{z'(z^{-1}(x))} \right| \cdot f^\infty_{\frac{1}{K}\mathbf{H}_1\mathbf{H}_1^H}(z^{-1}(x))=\frac{1}{\bar \mu}f_\mathrm{MP}\left(\frac{x-1}{\bar \mu}\right).
\label{eq: M aepdf}
\end{align}
\end{proof}
\end{lem}
\begin{thm}
\label{thm: eta transform M}
The inverse $\eta$-transform of $\mathbf{M}$ is given by \eqref{eq: inverse eta transform M}.

\begin{longequation*}[tp]
\begin{align}
\eta^{-1}_{\mathbf{M}}(x)={\frac {-x\bar \mu-\beta\,\bar \mu+\bar \mu-1+\sqrt {{x}^{2}{\bar \mu}^{2
}+2\,x\bar {\mu}^{2}\beta-2\,x\bar {\mu}^{2}-2\,x\bar \mu+{\beta}^{2}{\bar \mu
}^{2}-2\,\beta\,{\bar \mu}^{2}+2\,\beta\,\bar \mu+{\bar \mu}^{2}+2\,\bar \mu+1
}}{2x\bar \mu}}.
\label{eq: inverse eta transform M}
\end{align}
\hrule
\end{longequation*}
\begin{proof}
See Appendix \ref{app: 1}.\end{proof}
\end{thm}

\begin{thm}
\label{th: M aecdf}
The a.e.c.d.f. of $\mathbf{M}$ for \(\beta=1\) is given by:

\begin{equation}
F_{\mathbf{M}}(x)=\frac{\sqrt{\left( x-1\right)\left( 4\bar \mu-x+1\right)}-2\arcsin \left( {\frac {2\bar \mu-x+1}{ 2\bar \mu }} \right) \mu+\pi\bar \mu } {2{\pi }\bar \mu  }.
\end{equation}
\begin{proof}
The c.d.f. follows from eq.\eqref{eq: I plus MP aepdf} after integration for \(\beta=1\).
\end{proof}
\end{thm}
\begin{thm}
\label{thm: ieta transform K}
The inverse $\eta$-transform of $\mathbf{K}$ is given by:
\begin{equation}
\eta^{-1}_{\mathbf{K}}(x)=\Sigma_{\mathbf{\tilde N}}(x-1)\eta^{-1}_{\mathbf{M}}(x)
\label{eq: inverse eta transform K}
\end{equation}
\begin{proof}
Given the asymptotic freeness between deterministic matrix with bounded eigenvalues  $\mathbf{\tilde N}$ and unitarily invariant matrix $\mathbf M $, the $\Sigma$-transform of $\mathbf{K}$ is given by multiplicative free convolution:
\begin{align}
\Sigma_{\mathbf{K}}(x)&=\Sigma_{\mathbf{\tilde N}}(x)\Sigma_{\mathbf{M}}(x)\nonumber\stackrel{_{(a)}}{\Longleftrightarrow}\\
\left(-\frac{x+1}{x}\right)\eta^{-1}_{\mathbf{K}}(x+1)&=\Sigma_{\mathbf{\tilde N}}(x)\left(-\frac{x+1}{x}\right)\eta^{-1}_{\mathbf{M}}(x+1)\nonumber
\end{align}
where step $(a)$ combines Definition \ref{def: Sigma transform} and eq. \eqref{eq: sigma transform N}.
The variable substitution $y=x+1$ yields eq. \eqref{eq: inverse eta transform K}.
\end{proof}
\end{thm}

\subsection{Ill-conditioned Second Hop}
\label{sec: Ill-Conditioned Second Hop}

 Matrix \(\mathbf H_2\) is modelled as a deterministic matrix with power normalization \(\mathrm{tr}( \mathbf{H}_2^H\mathbf{H}_2 )=K\). Due to the lack of scattering in line-of-sight environments, this matrix may be ill-conditioned. The simplest model would be to assume a uniform distribution of eigenvalues with support \([\zeta^{-1}, \zeta]\) and condition number $\zeta^2$. The a.e.p.d.f. and transforms for
uniform eigenvalue distribution with variable condition number are given by:
\begin{align}
f^\infty_{\mathbf{\tilde N}}(x)&=\frac{\zeta}{\zeta^2-1}\mathbf{1}\left\{ \zeta^{-1}\ldots \zeta \right\}
\\{\eta_{\mathbf{\tilde  N}}(x)}&={\frac {\zeta\, \left( \ln  \left( \zeta \right) -\ln  \left( \zeta
+x \right) +\ln  \left( 1+x\zeta \right)  \right) }{ \left( {\zeta}^
{2}-1 \right) x}}
\\{\mathcal{S}_{\mathbf{\tilde N}}(x)}&={\frac {\zeta\, \left( \ln  \left( \zeta \right) -\ln  \left( x
\zeta-1 \right) +\ln  \left( x-\zeta \right)  \right) }{{\zeta}^{2}
-1}}.
\end{align}
However, this model results in exponential expressions for the \(R\)- and \(\Sigma\)-transforms which yields complex closed form expressions. To construct an analytically tractable problem, we consider the tilted semicircular law distribution which can accommodate a variable condition number and more importantly its \(\Sigma\)-transform is given by a first degree polynomial \cite{Mestre2003}. 
\begin{thm}
In the asymptotic regime preserving the power normalization, the tilted semicircular law converges to the following distribution: 
\begin{equation}
f^\infty_{\mathbf{\tilde N}}=\frac{2\zeta}{\pi\left(\zeta-1\right)^2 x^2}\sqrt{\left({\zeta x}-1\right)^{+}\left(1-\frac{x}{\zeta}\right)^+}
\label{eq: tsl}
\end{equation}
with support \([\zeta^{-1}, \zeta]  \). In this case, the  transforms
of the tilted semicircular law are given by:

\begin{equation}
{\eta_{\mathbf{\tilde N}}(x)}=\frac {1+2\zeta x+{\zeta }^{2}-2\sqrt {
 \zeta  \left( x+\zeta + \zeta{x}
^{2}+{\zeta }^{2}x \right) }}{\left(\zeta^{2}-1\right)^2}
\end{equation}
\begin{equation}
{\mathcal{S}_{\mathbf{\tilde N}}(x)}= \frac{-x+2\,\zeta -{\zeta }^{2}x+2\,\sqrt {{{ \zeta \left( -x+ \zeta  {x}^{2}+\zeta -{\zeta }
^{2}x \right) }}}}{{x}^{2} \left(\zeta^{2}-1\right)^2}
\end{equation}

\begin{equation}
{R_{\mathbf{\tilde N}}(x)}=\frac{2}{x}{\frac {\zeta-\sqrt {\zeta \left( \zeta +2\,
 \zeta x-x-{\zeta }^{2}x \right) } }{  \left(\zeta^{2}-1\right)^2}}
 \label{eq: tsl Sigma transform}
\end{equation}

\begin{equation}
{\Sigma_{\mathbf{\tilde N}}(x)}=1-\frac{\left(\zeta-1\right)^2}{4\zeta} x
\end{equation}

\begin{longequation*}[tp]
\begin{align}
\label{eq: V ill}
\scriptstyle{
{\mathcal{V}_{\mathbf{\tilde N}}(x)}= {\frac {2\,\sqrt {\zeta }\sqrt {x+\zeta }\sqrt {x \zeta 
+1}+\left(1+{\zeta }^{2}\right)\log  \left( 1+2\,x  \zeta   +{
\zeta }^{2}+2\,\sqrt {\zeta }\sqrt {x+\zeta }\sqrt {x  \zeta 
+1} \right) -2  \zeta  \log  \left( 2\,
\zeta +x+x{\zeta }^{2}+2\,\sqrt {\zeta }\sqrt {x+\zeta }\sqrt {x
\zeta  +1} \right) }{\left({\zeta }-1\right)^2}}
}\nonumber
\\\scriptstyle{
-2\frac{\,x \zeta  +\,
\zeta  -2\, \zeta   \,
\log  \left( 2 \right) -\,\zeta   \log  \left( \zeta 
 \right) +\,\left(1+{\zeta }^{2}\right)\log  \left( \zeta +1 \right)} {\left({\zeta }-1\right)^2}
}
\end{align}
\hrule
\end{longequation*}
\begin{proof}
The closed-form expressions for the transforms are derived by integrating over the aepdf \eqref{eq: tsl} using the definitions in app. \ref{app: 1}.
\end{proof}
\end{thm}

\begin{thm}
The capacity term \(\mathrm{C}_2\) is given in closed form using the Shannon transform:
\begin{equation}
\mathrm{C}_2=\mathcal{V}_{\mathbf{\tilde N}}\left( \nu \right)
\end{equation}
and in the low SNR regime:
\begin{equation}
\lim_{\nu\rightarrow0}\mathrm{C}_2=\frac {4\,\zeta -\left({\zeta }^{2}+4\zeta+1\right)\log  \left( 4\zeta \right) +4\left(\,{\zeta }^{2}
+1\right)\ln  \left( \zeta +1 \right)  }
{2\left({\zeta }-1\right)^2}.
\end{equation}
\begin{proof}
The first equation can be derived using eq. \eqref{eq: capacity aepdf 2} and def. \ref{def: Shannon transform}. As a result, \(\lim_{\nu\rightarrow0}\mathrm{C}_2=\mathcal{V}_{\mathbf{\tilde N}}\left( 0 \right)\). 
\end{proof}
\end{thm}

\begin{thm}
\label{thm: Stieltjes transform N}
The Stieltjes transform of $\mathbf{K}$ is given by the solution of the cubic polynomial in \eqref{eq: quartic}.
\begin{longequation*}[tp]
\begin{align}
\left( \left( \zeta- 1 \right) ^{4}\bar\mu\,{x}^{3}+ 4\,\zeta \left( \zeta- 1 \right) 
^{2}{\bar\mu}^{2}{x}^{2}
\right)&{{\mathcal{S}_{\mathbf{K}}(x)}}^{3}\nonumber
\\+  \left(\left( \zeta- 1 \right) ^{2} \left( 3\,{\zeta}^{2}+3+2\,\zeta \right) \bar\mu\,{x}^
{2}- \left(  \,\zeta \left( \zeta- 1 \right) ^{2}\beta- 2\zeta \left( 1-{
\zeta}^{2} \right)  \right) 4{\bar\mu}^{2}- 4\zeta \left( \zeta- 1 \right) ^{2}
\bar\mu\,x-(4\,{\bar\mu}{\zeta})^{2}\right)
 &{{\mathcal{S}_{\mathbf{K}}(x)}}^{
2}\nonumber
\\+  \left(\left( \zeta+ 1 \right) ^{2} \left( 3\,{\zeta}^{2}+3-2\,\zeta \right) \bar\mu\,x-
 4\left( \zeta+ 1\right) ^{2}\zeta \left(  \left( \beta-1 \right) {\bar\mu
}^{2}+\bar\mu \right) \right)
 &{\mathcal{S}_{\mathbf{K}}(x)}\nonumber
\\ \left( \zeta+ 1 \right) ^{4}\bar\mu&
\label{eq: quartic}
\end{align}
\hrule
\end{longequation*}
\begin{proof}
The first step is to substitute eq.
\eqref{eq: inverse eta transform M} and \eqref{eq: tsl Sigma transform} into \eqref{eq: inverse eta transform K}. Using prop. \ref{pro: Sigma n} and applying suitable change of variables:
\begin{equation}
x\eta^{-1}_{\mathbf{K}}\left(-x{\mathcal{S}_{\mathbf{K}}(x)}\right)+1=0.
\end{equation}
The final form of the polynomial is derived through algebraic calculations.
\end{proof}
\end{thm}

\begin{rem}
\label{rem: tilde K aepdf}
For \(M=K\), the eigenvalues of \(\mathbf K\) and $\mathbf {\tilde K}$ are identical. Thus, the a.e.p.d.f.
of $\mathbf {\tilde K}$ is given by eq. \eqref{eq: quartic} and Lemma \ref{pro: Stieltjes to aepdf} for \(\beta=1\).
\end{rem}
\begin{lem}
\label{pro: Stieltjes to aepdf M}
The quantity \(\mathrm{C}_1\) is given by eq. \eqref{eq: capacity aepdf 1},
where \(f^\infty_{\mathbf{K}}\) is given by lem. \ref{pro: Stieltjes to aepdf} and eq. \eqref{eq: quartic}.
\end{lem}

\begin{rem}
\label{rem: MMSE avg analytic}
The average MMSE \(\mathrm{mmse_{avg}}\) is given by eq. \eqref{eq: average MMSE} where \({F^{-1}_{\mathbf{\tilde M}}(x)}\) can be calculated using Theorem \ref{th: M aecdf} and \({F^{-1}_{\mathbf{\tilde K}}(x)}\) using integration and inversion over the a.e.p.d.f. in Remark \ref{rem: tilde K aepdf}.
\end{rem}
\subsection{Ill-conditioned Extreme \(\mu/\nu\) Limits }
\label{subsec: SNR limits}
\subsubsection{High-$\mu$, low-$\nu$ limit} 
In this regime, we consider the case of high FHP and low SHP for ill-conditioned second hop. Assuming \(\mu\rightarrow\infty, \nu\rightarrow0\) with constant \(\nu\mu\), the quantity \(\mathrm C\) can be written as:

\begin{align}
&\lim_{\stackrel{\mu\rightarrow\infty}{_{\nu\rightarrow0}}}\mathrm C=\lim_{\stackrel{\mu\rightarrow\infty}{_{\nu\rightarrow0}}}\mathrm{C}_1\nonumber
\\&=\frac{1}{K}\lim_{K,M,\stackrel{\mu\rightarrow\infty}{_{\nu\rightarrow0}}}\mathbb{E}\left[\log\det\left(\mathbf{I}+\nu\mathbf{H}_2^H\mathbf{H}_2\left(\mathbf{I}+\mu\mathbf{H}_\mathrm{1}\mathbf{H}_\mathrm{1}^H\right)\right)\right]\nonumber\\
&=\frac{1}{K}\lim_{K,M\rightarrow\infty}\mathbb{E}\left[\log\det\left(\mathbf{I}+\nu\mu\mathbf{\tilde N N}\right)\right]
\label{eq: scatter channel}
\end{align}
For ill-conditioned matices, the expression in \eqref{eq: scatter channel} is identical to the one derived in \cite{Mestre2003} for a single hop channel with one side correlation modeled according to the tilted semicircular law. As a result, it can be seen that investing on FHP and minimizing SHP results in a spatially correlated channel that does not suffer from noise amplification. In this case, high condition number entails high correlation for the equivalent channel. The free-probabilistic analysis of \eqref{eq: scatter channel} yields a closed-form expression  for \(\mathcal{S}_{\mathbf{\tilde NN}}\),  aepdf and capacity\cite[eq.(13-16)]{Mestre2003}.
In the same direction, the average MMSE in this regime can be simplified into: 
\begin{align}
&\mathrm{\lim_{\stackrel{\mu\rightarrow\infty}{_{\nu\rightarrow0}}}mmse_{avg}}
=\lim_{K,M,N\rightarrow\infty}\mathbb{E}\left[\frac{1}{M}\mathrm{tr}\left\{ \left(\mathbf{I}+\nu\mu\mathbf{\tilde N N}\right)^{-1}\right\}\right]=\eta_{\mathbf{\tilde N N}}(\nu\mu ).
\end{align}
Subsequently, using prop. \ref{pro: Sigma n} and \(\mathcal{S}_{\mathbf{\tilde NN}}\) from \cite{Mestre2003}, the following closed-form expression can be derived for \(\beta\leq1\):
\begin{align}
\eta_{\mathbf{\tilde N N}}(x)=&\scriptstyle{
\left( -{\zeta }^{2}-1+2\,x \zeta   -2\,\beta\,x
  \zeta   +2\,\sqrt {1+2\,x+2\,\beta\,x+{x}^{2}+2\,\beta
\,{x}^{2}+{\beta}^{2}{x}^{2}-\beta x \left({{4\,x \zeta   -\left({\zeta }-1\right)^{2}}}
 \right) }   \right)  \left( {\frac 
{  \left({\zeta }-1\right)^{2}-4\,x \zeta}{ \left(4x\zeta\right)^2 
  }} \right) }.
  \label{eq: Mestre eta}
\end{align}
\subsubsection{High-$\nu$ limit}
\label{subsec: high nu}
In this regime, we consider a capacity bound on symmetric systems for high SHP. Using Minkowski's inequality and \(\beta=\gamma=1\), the capacity can be lower bounded as:
\begin{align}
&\mathrm C\geq\log\left( 1+\nu\exp\left( \frac{1}{K}\mathbb{E}\left[\log\det\left(\mathbf{H}_{2}^H\mathbf{H}_{2}\left(\mathbf{I}+{\mu}\mathbf{H}_{1}\mathbf{H}_{1}^H\right)\right)\right] \right) \right)\nonumber\\&-\log\left( 1+\nu\exp\left( \frac{1}{K}\mathbb{E}\left[\log\det\left(\mathbf{H}_{2}\mathbf{H}^H_{2}\right)\right] \right) \right)\nonumber
\\&=\log\left( 1+\nu\exp\left( \frac{1}{K}\mathbb{E}\left[\log\det\left(\mathbf{\tilde N}\right)\right]+\frac{1}{K}\mathbb{E}\left[\log\det\left(\mathbf{M}\right)\right] \right) \right)\nonumber
\\&-\log\left( 1+\nu\exp\left( \frac{1}{K}\mathbb{E}\left[\log\det\left(\mathbf{\tilde N}\right)\right] \right) \right)\nonumber
\end{align}
In the high-$\nu$ limit, using \(\lim_{\nu\rightarrow\infty}\log( 1+\nu x)=\log( \nu x)\)
\begin{equation}
\lim_{\nu\rightarrow\infty}C\geq\frac{1}{K}\mathbb{E}\left[\log\det\left(\mathbf{M}\right)\right]=\mathcal{V}_{\mathbf{N}}\left (\mu,\beta\right)\rightarrow\mathcal{V}_{\mathrm{MP}}\left (\bar \mu,\beta\right)
\end{equation}
where \(\mathcal{V}_{\mathrm{MP}}\) is given by def. \ref{df: MP law}. As a result, it is shown that for high SHP the performance becomes independent of the characteristics of the second hop and it is entirely governed by the first hop. More specifically, this tight bound corresponds to the capacity of the first hop, which acts like bottleneck in this regime.

Similarly, it can be seen  that this result also applies for the average MMSE:
\begin{align}
\lim_{\nu\rightarrow\infty}\mathrm{mmse_{avg}}&=\mathbb{E}\left[\frac{1}{M}\mathrm{tr}\left\{ \left(\mathbf{I}+{\mu}\mathbf{H}_{1}\mathbf{H}_{1}^H\right)^{-1} \right\}\right]\nonumber
\\&=\eta_{\mathbf{N}}\left (\mu,\beta\right)\rightarrow\eta_{\mathrm{MP}}\left (\bar\mu,\beta\right).
\end{align}

\subsection{Rank-deficient Second Hop}
\label{sec: rank-deficient second hop}
Rank deficiency is an extreme form of ill-conditioning where zero eigenvalues appear. In some problems, rank deficient matrices are used to approximate ill-conditioned ones by substituting infidecimal eigenvalues with zero. In order to model rank deficiency, we consider a \(K\times K\) channel matrix \(\mathbf H_2\) with \(\mathrm{tr}\{\mathbf H_2\mathbf H_2^H\}=K\) and \(\mathrm{rank}\{\mathbf H_2\mathbf H_2^H\}=\alpha K\) where \(\alpha\in(0,1)\) is the normalized rank, namely the matrix rank normalized
by the matrix dimension. By varying \(\alpha\) from zero to unity, we can recover unit- to full-rank matrices.

\begin{thm}
\label{thm: rank-deficient}
In the asymptotic regime, the capacity converges to 

\begin{equation}
\label{eq: C rank}
\mathrm{C}\rightarrow\alpha\mathcal{V}_{\mathrm{MP}}\left(\alpha\frac{\bar\mu\nu}{\nu+\alpha},\frac{\beta}{\alpha}\right).
\end{equation}
The a.e.p.d.f. of matrix $\mathbf K$ follows a scaled version of the MP law:

\begin{equation}
f^\infty_{\mathbf{K}}=\frac{\alpha}{\bar\mu}f_{\mathrm{MP}}\left(\frac{{\alpha}{x-1}}{\alpha\bar\mu},\frac{\beta}{\alpha}\right).
\end{equation}
\begin{proof}
See Appendix \ref{app: C}.
\end{proof}
\end{thm}
\begin{rem}
For rank-deficient second-hop, the MMSE performance degrades rapidly since the equivalent receive dimensions are fewer that the number of users. As a result, the MMSE receiver could only be used if the channel rank is larger that the number of served users \(\alpha \geq \beta\). 
\end{rem}

\section{Numerical Results}
\label{sec: numerical results}
In order to verify the accuracy of the derived closed-form expressions and gain some insights on the system performance of the considered  model, a number of numerical results are presented in this section.  

\subsection{A.e.p.d.f. Results}
\begin{figure}
   \centering
  \includegraphics[width=0.45\textwidth]{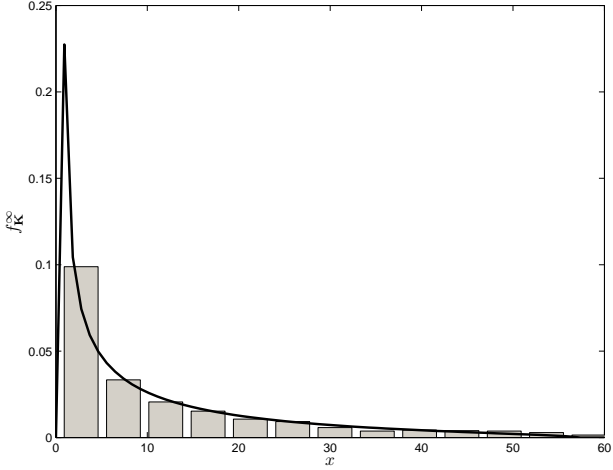}
  \caption{A.e.p.d.f. plots of matrix $\mathbf{K}$. Parameters: $\beta=1,\nu=\mu=10dB$. The solid analytic curves follow tightly the simulation-generated bars.}
  \label{fig:1a}
\end{figure}
\begin{figure}
   \centering
  \includegraphics[width=0.45\textwidth]{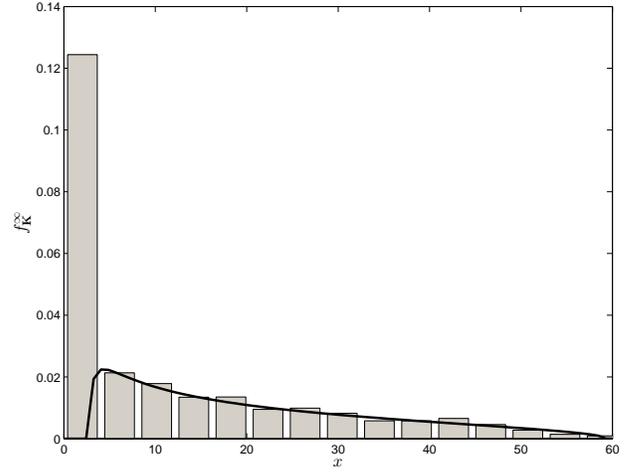}
  \caption{A.e.p.d.f. plots of matrix $\mathbf{M}$. Parameters: $\beta=1,\nu=\mu=10dB$. The solid analytic curves follow tightly the simulation-generated bars.}
\label{fig:1b}
\end{figure}
The accuracy of the derived closed-form expressions for the a.e.p.d.f. of matrices $\mathbf{K,M}$ is depicted in Figures \ref{fig:1a} and \ref{fig:1b} for ill-conditioned  second hop. The solid line in subfigure \ref{fig:1a} is drawn using Theorem \ref{thm: Stieltjes transform N} in combination with lem. \ref{pro: Stieltjes to aepdf} , in subfigure \ref{fig:1b} using lem. \eqref{lem: M aepdf}. The histograms denote the p.d.f. of matrices $\mathbf{K},\mathbf{M}$ calculated numerically based on Monte Carlo simulations for \(K=10\).  It can be seen that there is a perfect agreement \footnote{The far left bar in subfig. \ref{fig:1b} corresponds to the zero-eigevalues Dirac delta $\delta(x)$ which occurs due to rank deficiency (e.g. see eq. \eqref{eq: MP aepdf}).} between the two sets of results which verifies our analytic results.

\subsection{Capacity Results}
\begin{figure}
        \centering
                \includegraphics[width=0.45\textwidth]{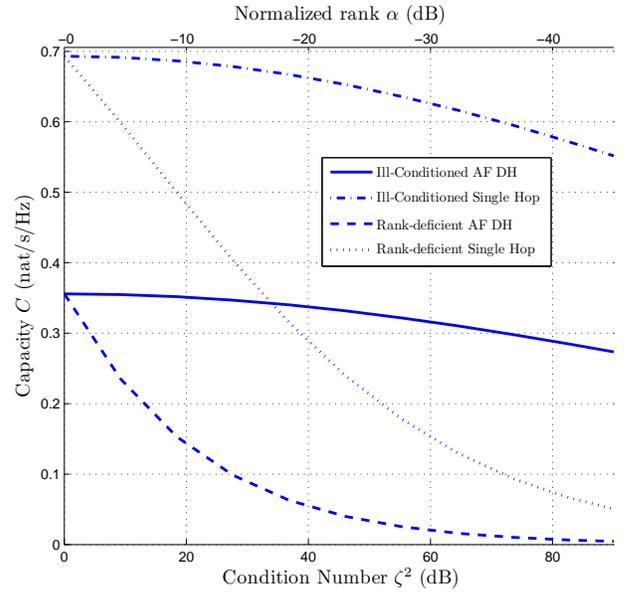}
        \caption{Per-antenna capacity scaling vs. condition number $\zeta^{2}$ and normalized rank $\alpha$ in dBs. Parameters: $\mu=\nu=\beta=1$.}
        \label{fig: capacity scaling b}
\end{figure}

\begin{figure}
        \centering
                \includegraphics[width=0.45\textwidth]{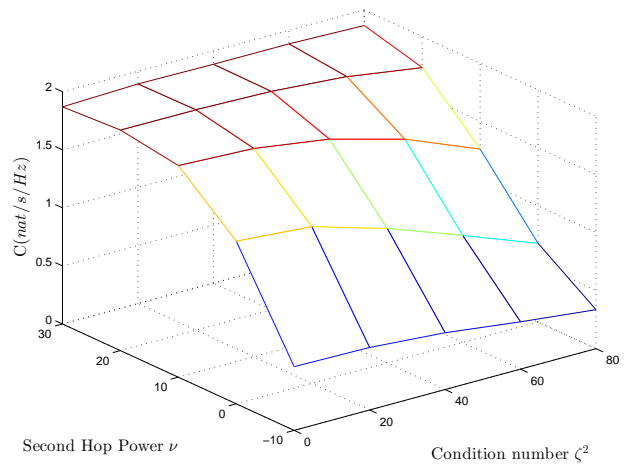}
        \caption{Per-antenna capacity scaling vs. condition number $\zeta^2$ and second hop power $\nu$ in dBs. Parameters: $\mu=10dB, \beta=1$.}
        \label{fig: capacity scaling SNR}
\end{figure}
Figures \ref{fig: capacity scaling b} and  \ref{fig: capacity scaling SNR} depict the effect of  condition number $\zeta^{2}$ and normalized rank $\alpha$ on the per-antenna channel capacity $\mathrm{C}$ of the DH AF SIMO MAC and the per-antenna channel capacity \(\mathrm C_2\) which corresponds to an ill conditioned or rank-deficient single hop SIMO MAC respectively. The analytic solid curves are plotted using a) eq. \eqref{eq: capacity aepdf 1} and eq. \eqref{eq: capacity aepdf 2} for the ill-conditioned
DH AF, b)
eq. \eqref{eq: C rank} for the rank deficient
DH AF, c) eq. \eqref{eq: V ill} for the ill-conditioned single hop and d) eq. \eqref{eq: V rank} for the rank deficient single hop.
It can be seen that the performance 
 degrades much more steeply with normalized rank than condition number in all cases. Especially for the DH AF, it can be observed that rank deficiency is detrimental and quickly drives capacity to zero due to rank loss. On the other hand, the degradation with condition number is much smoother since the second hop channel matrix \(\mathbf H_2\) is still full-rank. 

In addition, the per-channel capacity \(\mathrm C\) is plotted versus the second hop power \(\nu\) and condition number $\zeta^2$. As it can be seen, it is possible to recover part of the lost performance due to ill-conditioning by increasing the amplification level \(\nu\).

\subsection{MMSE Results}
\begin{figure}
        \centering
                \includegraphics[width=0.45\textwidth]{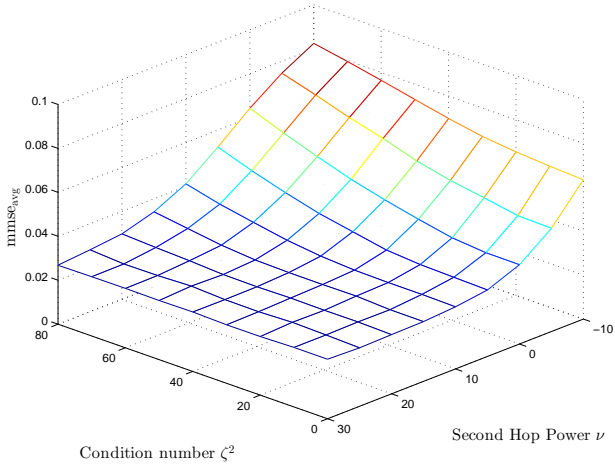}
        \caption{Average MMSE scaling vs. condition number $\zeta^2$ and second hop power $\nu$ in dBs. Parameters: $\mu=10dB, \beta=1$. For high amplification, first hop performance acts as bottleneck. }
        \label{fig: average mmse scaling SNR}
\end{figure}


\begin{figure}
  \centering
  \includegraphics[width=0.45\textwidth]{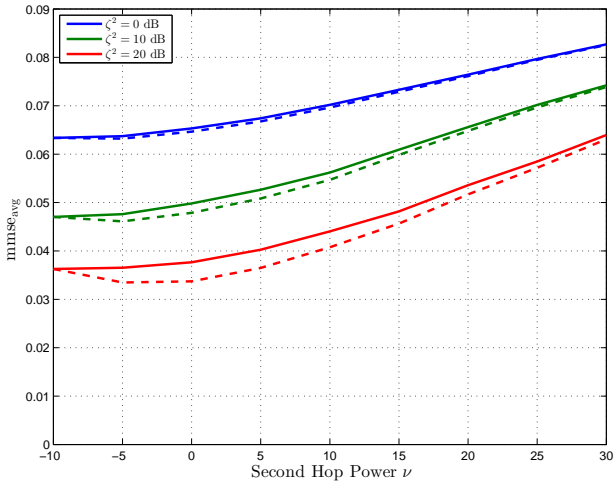}
  \caption{Average MMSE performance (solid line) and proposed lower bound (dashed line) vs second hop power \(\nu\). Parameters: $\mu=10dB$}
\label{fig:4a}
\end{figure}
\begin{figure}
  \centering
   \includegraphics[width=0.45\textwidth]{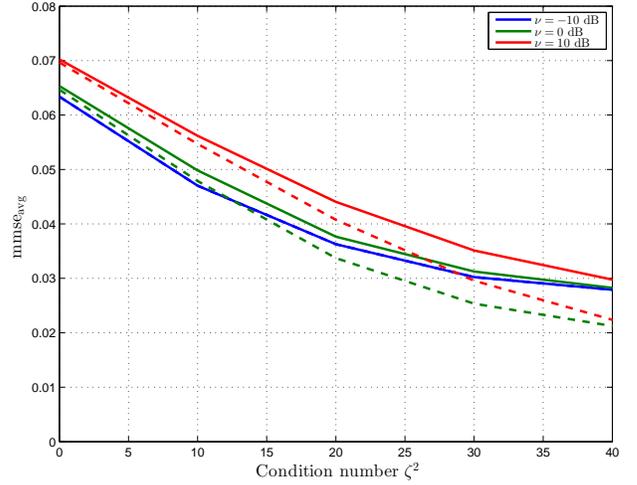}
  \caption{Average MMSE performance (solid line) and proposed lower bound (dashed line) vs condition number \(\zeta^2\). Parameters: $\mu=10dB$}
\label{fig:4b}
\end{figure}
Figure \ref{fig: average mmse scaling SNR} depicts the effect of condition number $\zeta^2$ and second hop power $\nu$ on the average MMSE.
As expected,  the average MMSE increases with  $\zeta^2$ but decreases with \(\nu\). It can be seen that performance can be improved using stronger amplification but for high \(\nu\) there is a saturation threshold which is governed by the first hop performance as described in sec. \ref{subsec: high nu}. 
Figures \ref{fig:4a} and  \ref{fig:4b} depicts the accuracy of the proposed lower bound. The solid  plots were calculated through Monte Carlo simulations of eq. \eqref{eq: mmse avg simulation}, whereas the dashed  plots represent our lower bound which was calculated using Remark \ref{rem: MMSE avg analytic}.  It can be seen that the proposed bound is tight  for low values of \(\zeta^2\), but it progressively diverges as \(\nu\) and $\zeta^2$ grow large. 

\subsection{Comparison}
\label{subsec: comparison}
\begin{figure}
        \centering
                \includegraphics[width=0.45\textwidth]{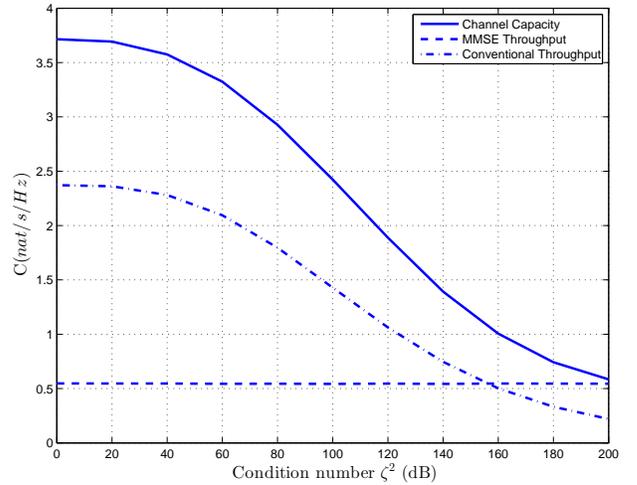}
        \caption{Throughput comparison between proposed and conventional system vs. condition number $\zeta^2$ in dBs. Parameters: $\mu=\nu=10dB, \beta=1$. The proposed system is preferable for condition numbers up to \(120\) dBs. }
        \label{fig: comparison}
\end{figure}
In this section, the performance of the proposed system is compared to the conventional system (as described in section \ref{subsec: conventional}) by fixing the user and BS power at \(10\) dBs. As it can be seen in Fig. \ref{fig: comparison}, while the condition number increases, the performance of the proposed system degrades and even falls below conventional performance for extremely ill-conditioned BS-CP channels. There are two crossing points in \(160\) and \(200\) dBs for the MMSE throughput
and channel capacity respectively.
However, a two-fold performance gain can still be harnessed for condition numbers up to \(120\) dBs for MMSE receiver and up to \(160\) dBs for optimal receiver.

\section{Conclusion}
\label{sec: conclusion}
In this paper, we have investigated the performance of
BS cooperation scenario with microwave backhauling to a
CP, where multiple users and BSs share the same channel
resources. The user signals are forwarded by the BSs to an
antenna array connected to a CP which is responsible for multiuser joint processing. This system has been modelled as a DH AF SIMO MAC with a ill-conditioned or rank-deficient second hop due to lack of scattering in line-of-sight environments. Its performance in terms of channel capacity and MMSE performance has been analysed through a large-system free-probabilistic analysis. It can be concluded that performance degrades much more gracefully with condition number than with loss of rank. As a result, a performance gain can be achieved compared to conventional resource partitioning even for highly ill-conditioned second hop. Furthermore, performance degradation due to ill conditioning can be compensated through stronger amplification at BS-side until it reaches the first hop performance in the high amplification limit. 
\appendices
\section{Random Matrix Theory Preliminaries}
\label{ap: preliminaries}
Let $f_\mathbf{X}(x)$ be the eigenvalue probability distribution function of a matrix \(\mathbf X.\)
\begin{df}
\label{def: Shannon transform}
The Shannon transform of a positive semidefinite matrix \(\mathbf X\) is defined as
\begin{equation}
\mathcal{V}_{\mathbf X}\left(\gamma\right)=\int_0^\infty\log\left({1+\gamma x}\right)f_\mathbf{X}(x)dx.
\end{equation}
\end{df}
\begin{df}
\label{def: n transform}
The $\eta$-transform of a positive semidefinite matrix \(\mathbf X\) is defined as
\begin{equation}
\eta_{\mathbf X}\left(\gamma\right)=\int_0^\infty\frac{1}{1+\gamma x}f_\mathbf{X}(x)dx.
\label{eq: eta def}
\end{equation}
\end{df}
\begin{df}
\label{def: Sigma transform}
The $\Sigma$-transform of a positive semidefinite matrix \(\mathbf X\) is defined as\begin{equation}
\Sigma_{\mathbf{X}}(x)=-\frac{x+1}{x}\eta^{-1}_{\mathbf{X}}(x+1).
\end{equation}
\end{df}
\begin{pro}
\label{pro: Sigma n}
The Stieltjes-transform of a positive semidefinite matrix \(\mathbf X\) can be derived by its $\eta$-transform using
\begin{equation}
\mathcal{S}_{\mathbf{X}}(x)=-\frac{\eta_{\mathbf{X}}(-1/x)}{x}.
\end{equation}
\end{pro}
\begin{lem}
\label{pro: Stieltjes to aepdf}
The a.e.p.d.f. of $\mathbf{X}$ is obtained by determining the imaginary part of the Stieltjes transform \(\mathcal{S}\) for real arguments
\begin{equation}
f^{\infty}_\mathbf{X}(x)=\lim_{y\rightarrow0^+}\frac{1}{\pi}\mathfrak{I}\left\{ \mathcal{S}_\mathbf{X}(x+\mathrm{j}y)\ \right\}.
\label{eq: limiting eigenvalue pdf}
\end{equation}
\end{lem}

\section{Proof of Theorem \ref{thm: eta transform M}}
\label{app: 1}

Starting from eq. \eqref{eq: M aepdf} and following def. \ref{def: n transform}:
\begin{align}
\eta_\mathbf{M}(\psi)&=\int_{-\infty}^{{+\infty}}\frac{1}{1+\psi x}f^\infty_\mathbf{M}(x)dx\nonumber\\
&=\frac{1}{\gamma}\int_{-\infty}^{{+\infty}}\frac{1}{1+\psi x}{f}_\mathbf{\tilde N}\left(\frac{x-1}{\gamma}\right)dx\nonumber\\
&\stackrel{_{(a)}}{=}\gamma\frac{1}{4i\pi}\oint_{\left|\zeta\right|=1}\frac{(\zeta^2-1)^2}{\zeta((1+\beta)\zeta+\sqrt{\beta}(\zeta^2+1))(\zeta(1+\psi(1+\gamma+\gamma\beta))+\sqrt{\beta}\psi\gamma(\zeta^2+1))}d\zeta.
\label{eq: eta derivation}
\end{align}
Step $(a)$ requires the variable substitutions $x=w\gamma+1$, $dx=\gamma dw$, followed by $w=1+\beta+2\sqrt{\beta}\cos \omega$, $dw=2\sqrt{\beta}(-\sin\omega)d\omega$ and finally $\zeta=e^{i\omega}$,  $d\zeta=i\zeta d\omega$\cite{Bai09}.
Subsequently, a Cauchy integration is performed by calculating the poles $\zeta_i$ and residues $\rho_i$ of eq. \eqref{eq: eta derivation}:
\begin{align}
\zeta_0&=0,\nonumber\\
\zeta_{1,2}&=\frac{-(1+\beta)\pm(1-\beta)}{2\sqrt{\beta}},\nonumber\\
\zeta_{3,4}&={\frac {-1-\psi\,\gamma-\psi\,\beta\,\gamma-\pm \left( \psi \sqrt{1+2\,\psi+2\,\psi\,\gamma+2\,\psi\,\beta\,\gamma+{
\psi}^{2}+2\,{\psi}^{2}\gamma+2\,{\psi}^{2}\beta\,\gamma+{\psi}^{2}{
\gamma}^{2}-2\,\beta\,{\psi}^{2}{\gamma}^{2}+{\psi}^{2}{\beta}^{2}{
\gamma}^{2}} \right) }{ 2\sqrt{\beta}\psi\,\gamma}}
\nonumber
\label{eq:}
\end{align}
Using the residues which are located within the unit disk, the Cauchy integration yields:
\begin{equation}
\eta_\mathbf{M}(\psi)=-\frac{\beta}{2}(\rho_0+\rho_2+\rho_4)\nonumber
\end{equation}
Inversion yields eq. \eqref{eq: inverse eta transform M}.

\section{Proof of Theorem \ref{thm: rank-deficient}}
\label{app: C}

The components of eq. \eqref{eq: generic capacity} can be written  as: 
\begin{align}
\mathrm{C}_1&=\frac{1}{K}\lim_{K,M\rightarrow\infty}\mathbb{E}\left[\log\det\left(\mathbf{I}_K+\nu\mathbf{H}_2^H\mathbf{H}_2\left(\mathbf{I}_K+\mu\mathbf{H}_\mathrm{1}\mathbf{H}_\mathrm{1}^H\right)\right)\right]\nonumber\\
&=\frac{1}{K}\lim_{K,M\rightarrow\infty}\mathbb{E}\left[\log\det\left(\mathbf{I}_K+\frac{\nu}{\alpha}\mathbf{D}\left(\mathbf{I}_K+\mu\mathbf{H}_\mathrm{1}\mathbf{H}_\mathrm{1}^H\right)\right)\right]\nonumber\\
&=\frac{1}{K}\lim_{K,M\rightarrow\infty}\mathbb{E}\left[\log\det\left(\mathbf{I}_{\alpha K}+\frac{\nu}{\alpha}\left(\mathbf{I}_{\alpha K}+\mu\mathbf{\bar H}_\mathrm{1}\mathbf{\bar H}_\mathrm{1}^H\right)\right)\right]\nonumber\\
&=\alpha \log\left( 1+\frac{\nu}{\alpha} \right)\nonumber\\&+\frac{1}{K}\lim_{K,M\rightarrow\infty}\mathbb{E}\left[\log\det\left(\mathbf{I}_{\alpha K}+\frac{aK\mu\nu}{\nu+\alpha}\frac{\mathbf{\bar H}_\mathrm{1}\mathbf{\bar H}_\mathrm{1}^H}{\alpha K}\right)\right]\nonumber\\
&\rightarrow\alpha \log\left( 1+\frac{\nu}{\alpha} \right)+\alpha\mathcal{V}_{\mathrm{MP}}\left(\alpha\frac{\bar\mu\nu}{\nu+\alpha},\frac{\beta}{\alpha}\right),
\end{align}
\begin{align}
\label{eq: V rank}
\mathrm{C}_2&=\frac{1}{K}\lim_{K\rightarrow\infty}\mathbb{E}\left[\log\det\left(\mathbf{I}_K+\nu\mathbf{H}_2^H\mathbf{H}_2\right)\right]\nonumber\\
&=\frac{1}{K}\lim_{K\rightarrow\infty}\mathbb{E}\left[\log\det\left(\mathbf{I}_{K}+\frac{\nu}{\alpha}\mathbf{D}\right)\right]\nonumber\\
&=\frac{1}{K}\lim_{K\rightarrow\infty}\mathbb{E}\left[\log\det\left(\mathbf{I}_{\alpha K}\left(1+\frac{\nu}{\alpha}\right)\right)\right]\nonumber\\
&\rightarrow\alpha\log\left(1+\frac{\nu}{\alpha}\right),
\end{align}
where \(\mathbf D\) is a \(K\times K\) zero matrix with \(\alpha K\) ones across its diagonal and \(\mathbf{\bar H}_1\) is a \( \alpha K\times M\) submatrix  of \(\mathbf{H}_1\).
Substraction yields the capacity expression. The aepdf follows from the equivalent matrix \(\mathbf K\):
\begin{equation}
\mathbf K=\frac{1}{\alpha}\left(\mathbf{I}_{\alpha K}+aK\mu\frac{\mathbf{\bar H}_\mathrm{1}\mathbf{\bar H}_\mathrm{1}^H}{aK}\right).
\end{equation}



\ifCLASSOPTIONcaptionsoff
  \newpage
\fi



\bibliographystyle{IEEEtran}
\bibliography{IEEEabrv,references,journals,books,conferences,thesis}

\end{document}